\documentclass[a4paper,11pt]{article}
\pdfoutput=1 

\usepackage{jheppub} 

\usepackage[T1]{fontenc} 
\usepackage{dsfont}
\usepackage{caption}
\usepackage{subcaption}
\newsavebox{\imagebox}

\allowdisplaybreaks
\def\myfloor#1{\left\lfloor {#1} \right \rfloor}
\newcommand{\CC}{C\nolinebreak\hspace{-.05em}\raisebox{.4ex}{\tiny\bf +}\nolinebreak\hspace{-.10em}\raisebox{.4ex}{\tiny\bf +}}

\title{\boldmath A holographic inequality for $N=7$ regions}

\author[a]{Bart{\l}omiej Czech and Yunfei Wang}
\affiliation[a]{Institute for Advanced Study, Tsinghua University, Beijing 100084, China}

\emailAdd{bartlomiej.czech@gmail.com, yunfeiwang0214@gmail.com}

\abstract{In holographic duality, boundary states that have semiclassical bulk duals must obey inequalities, which bound their subsystems' von Neumann entropies. Hitherto known inequalities constrain entropies of reduced states on up to $N=5$ disjoint subsystems. Here we report one new such inequality, which involves $N=7$ disjoint regions. Our work supports a recent conjecture on the structure of holographic inequalities, which predicted the existence and schematic form of the new inequality. We explain the logic and educated guesses by which we arrived at the inequality, and comment on the feasibility of employing similar tactics in a more exhaustive search.}

\begin{document} 
\maketitle
\flushbottom

\section{Introduction and review}

The Ryu-Takayanagi proposal \cite{rt} posits that in holographic duality (AdS/CFT correspondence \cite{adscft}) the von Neumann entropies of CFT regions are encoded in the bulk AdS geometry as areas of extremal surfaces. This fact imposes constraints on von Neumann entropies of boundary (CFT) states, which are dual to semiclassical bulk (AdS) geometries. The present paper reports one new such constraint. The new constraint---inequality~(\ref{ourineq})---is significant because it confirms a recent conjecture about an infinite class of constraints, which are characterized by a highly regular structure and a crisp physical interpretation \cite{withsirui} (see also \cite{cuencafadel}). 

The earliest-known constraint on entanglement entropies of states with semiclassical bulk duals is the \emph{monogamy of mutual information} \cite{refmmi}:
\begin{equation}
S_{AB} + S_{BC} + S_{CA} \geq S_A + S_B + S_C + S_{ABC}
\label{defmmi}
\end{equation}
Here $S_X$ is the von Neumann entropy of a boundary region $X$, and union signs are dropped, i.e. $AB \equiv A \cup B$. Since then, several other inequalities have been proven; see \cite{n5cuenca} for a complete list of currently known inequalities. This list comprises several inequalities on $N=5$ regions (usually denoted $A, B, C, D, E$), and one infinite class of inequalities that work for every $N$ \cite{hec}. The inequalities are said to define the \emph{holographic entropy cone} \cite{hec} because every inequality divides the vector space of potential entropy assignments (entropy space) into one allowed and one disallowed region, separated by a hyperplane where the inequality is saturated. A collection of hyperplanes (inequalities) defines a convex cone of allowed entropy assignments. 

Reference~\cite{hec} established an algorithmic technique for proving candidate holographic inequalities. It involves finding a so-called contraction map, which is a weak homomorphism of hypercubes whose dimensions are labeled by terms on the left and right hand side of the inequality, respectively.\footnote{In writing `left hand side' and `right hand side,' we are sticking to the convention of writing the inequality as `LHS $\geq$ RHS,' where all terms appear with positive coefficients. This convention is applied in (\ref{defmmi}) and throughout this paper.} A computer program, which searches for contraction maps, is publicly available \cite{michael}. In this paper, we assume that the reader is familiar with how contraction maps work,\footnote{For a pedagogical review, see \cite{myreview}.} or at least that she is willing to accept a computer-generated contraction map as proof of an inequality. Reference~\cite{hec} also established a method to confirm that an already-proven inequality is the tightest possible, i.e. that no terms can be added to the right hand side or subtracted from the left without falsifying the inequality.  That task involves certain graph models, which we briefly review in Section~\ref{sec:hyper}.  

Qualitatively speaking, the inequalities are difficult to interpret. Two exceptions include the monogamy of mutual information~(\ref{defmmi}), which was interpreted in \cite{mmipt} as quantifying perfect tensor-like entanglement of a four-partite state, and the infinite class of inequalities reported in \cite{hec}, which discretize a measure of entanglement called differential entropy \cite{diffent}. For other attempts to interpret known inequalities and/or identify future ones, see \cite{arrangement, kbasis, superbalance, sacone}. Moreover, known facts about the holographic entropy cone are subject to several important provisos. First, proofs by contraction \cite{hec} assume that the bulk geometry has time reflection symmetry. Therefore, it is logically possible that states and geometries with arbitrary time dependence might violate the inequalities listed in \cite{n5cuenca}.\footnote{Reference~\cite{withxi} proved that any inequality established by the contraction method remains valid for arbitrary states in the AdS$_3$/CFT$_2$ correspondence. Therefore, this hypothetical scenario would require a higher-dimensional setup, with at least four bulk dimensions. There is ongoing work, which aims to eliminate this logical possibility \cite{mattontime}.} Second, the inequalities really constrain the geometric (area-like) contributions to boundary von Neumann entropies, and ignore bulk entanglement. The latter term, which is part and parcel of the quantum-corrected Ryu-Takayanagi proposal \cite{qes, qes2}, is typically suppressed by the small holographic parameter $1/N$ \cite{flm}. In recent years, however, setups where area terms and bulk entanglement terms compete---dubbed \emph{islands}---have been used to explain the black hole information paradox \cite{islands1, islands2, islands3}. Although island-like setups do satisfy (indeed, saturate \cite{withsirui}) holographic entropy inequalities, they make the inequalities conceptually harder to interpret. This issue was discussed in \cite{qescone}. 

\subsection{Symmetrized inequalities and the HCAE conjecture}
\label{sec:hcae}
To make progress in studying the holographic entropy cone, Reference~\cite{withsirui} (see also \cite{cuencafadel}) proposed to study a simpler but more doable problem. The idea is to study inequalities, which bound \emph{averages} of von Neumann entropies of $p$-partite regions. For example, in the $N=3$ context we have two such quantities:
\begin{align}
S^1 & = 
\big( S_A + S_B + S_C + S_{O}\big)/4\\
S^2 & = 
\big( S_{AB} + S_{BC} + S_{CA} + S_{AO} + S_{BO} + S_{CO} \big) / 6
\end{align}
These expressions assume that the $3$-partite state on named regions $A,B, C$ is purified by a fourth region $O$. The averages $S^p$ are taken over all $p$-partite regions, also those which involve the purifier. As such, they form the complete set of invariants of the symmetric group $S_{4}$, which permutes all the regions, including the purifier. For this reason, constraints on quantities $S^p$ are \emph{symmetrized inequalities}. 

At general $N$, we will likewise envoke an $(N+1)^{\rm st}$ purifying region and treat it on equal footing with the $N$ named regions. Because complementary regions of a pure state have equal entropies, in the $N$-region context we have 
\begin{equation}
S^p = S^{N+1-p} 
\label{complementsid}
\end{equation}
Therefore, the complete set of entropy averages $S^p$ in the $N$-region context is labeled by $1 \leq p \leq \myfloor{(N+1)/2}$, where $\myfloor{\ldots}$ is the floor function.

An example inequality, which bounds the averages $S^p$, is the monogamy of mutual information. Written in terms of $S^p$, inequality~(\ref{defmmi}) is simply:
\begin{equation}
3S^2 \geq 4 S^1 \label{mmirewrite}
\end{equation}
Going to higher $N$, Reference~\cite{withsirui} conjectured the following infinite class of inequalities, parameterized by $p$:
\begin{equation}
\frac{2 S^p}{p} \geq \frac{S^{p-1}}{p-1} + \frac{S^{p+1}}{p+1}
\qquad\qquad {\rm where}~~2 \leq p \leq \myfloor{(N+1)/2}
\label{conjecture}
\end{equation}
Note that in the context of $N=3$ regions ($A, B, C$), the monogamy of mutual information~(\ref{mmirewrite}) is inequality~(\ref{conjecture}) for $p=2$. The $N=3$ assumption is used for setting $S^1 = S^3$ by equation~(\ref{complementsid}). 

The conjectured inequalities~(\ref{conjecture}), if true, enjoy a wealth of fascinating and intuitive properties:
\begin{itemize}
\item Complemented by $2S^1 \geq S^2$, inequalities~(\ref{conjecture}) form the complete and tightest possible set of inequalities, which bound $S^p$. They define the \emph{holographic cone of average entropies} (HCAE).
\item They collectively define a pattern of most efficient (average) purification of a $p$-partite region with the addition of one extra region. This can be expressed as the bound, which follows from (\ref{conjecture}):
\begin{equation}
\frac{S^{p+1}}{S^p} \geq \frac{(p+1)(N-p)}{p(N+1-p)}
\label{lowerbound}
\end{equation}
The most efficient pattern of purification is when inequality~(\ref{lowerbound}) is saturated.
\item Saturating~(\ref{lowerbound}) sets $S^{p'} \propto p' (N+1-p')$ for all $p' \geq p$. The same behavior of $S^{p'}$ characterizes a collection of EPR pairs. In other words, the most efficient purification of $p$-partite regions has the same (average) von Neumann entropies of $(p'\geq p)$-partite regions as would a collection of EPR pairs.
\item When a maximal number of inequalities are collectively saturated, we get an \emph{extreme ray} of the cone. The extreme rays of HCAE correspond to stages of evaporation of an old black hole, as characterized by Page \cite{page1, page2} and later in the islands proposal \cite{islands1, islands2, islands3}. This is the physical interpretation of inequalities~(\ref{conjecture}), which we advertised in the opening paragraph. For more details, see \cite{withsirui}.
\end{itemize}
In this paper, we refer to inequalities~(\ref{conjecture}) collectively as \emph{the HCAE conjecture}.

One challenge with inequalities~(\ref{conjecture}) is that they cannot be proven using computer programs \cite{michael}, which search for contraction maps. The limitation is not fundamental but practical. A contraction is a weak homomorphism from $(\mathbb{Z}_2)^L$ to $(\mathbb{Z}_2)^R$, where $L$ and $R$ are the numbers of terms on the left/right hand side. If we take $p = \myfloor{(N+1)/2}$, inequality~(\ref{conjecture}) has $L \geq \frac{1}{2}\binom{2p}{p}$ types of terms on the left hand side and $R \geq \binom{2p}{p-1}$ types of terms on the right. We write $\geq$ instead of $=$ because in practice the incommensurate coefficients $2/p$, $1/(p-1)$ and $1/(p+1)$ make the task of finding a contraction even more computationally challenging.\footnote{Identical terms such as $2S_X$ may need to be treated as independent (that is, $S_X + S_X$) for the purpose of building a contraction. In that instance parameters $L$ and $R$ become greater than their na{\"\i}ve values.} A brute force search for a contraction map that directly establishes~(\ref{conjecture}) might be marginally feasible for $p=3$, but not beyond.

Efforts to establish~(\ref{conjecture}) are ongoing \cite{selfref}. The obvious method is to construct a contraction map by hand, for arbitrary $p$ and $N$. One result of the present paper is to prove inequality~(\ref{conjecture}) for $p=4$ (and arbitrary $N \geq 7$).

\paragraph{Symmetrizing non-symmetric inequalities}
Every detailed inequality---that is, one bounding specific entropies like $S_A$ and not merely their averages $S^p$---gives rise to a symmetrized inequality. When we add up all permutation images of a given detailed inequality, by definition we obtain an inequality that only involves permutation invariants, i.e. the quantities $S^p$. The coefficients, with which the $S^p$ enter the symmetrized inequality, are simply copied from the parent inequality. This is because adding permutation-equivalent inequalities does not change the proportion, with which various $p$-partite terms appear.

In summary, symmetrizing a detailed inequality is carried out by the simple procedure:
\begin{itemize}
\item Replace every $p$-partite term with $S^p$.
\end{itemize}
To clarify this, we give two examples---subadditivity and the $N=5$ cyclic inequality \cite{hec}:
\begin{equation}
S_A + S_B \geq S_{AB} \quad \longrightarrow \quad 2S^1 \geq S^2
\end{equation}
\begin{align}
S_{ABC} + S_{BCD} + S_{CDE} + S_{DEA} + S_{EAB} 
& \geq S_{AB} + S_{BC} + S_{CD} + S_{DE} + S_{EA} + S_{ABCDE} \nonumber \\ 
\longrightarrow \quad 5 S^3 & \geq 5 S^2 + S^5
\end{align}
When working with a fixed number of regions $N$, we can further rewrite these symmetrizations using~(\ref{complementsid}). For instance, in the second example, we can also write $5S^3 \geq 5S^2 + S^1$ in the $N=5$ context or $5S^3 \geq 5S^2 + S^4$ in the $N=8$ context.

Inequalities known thus far have the following symmetrizations:
\begin{align*}
i_1 & \qquad \longrightarrow \qquad 2S^1 \geq S^2 \\
i_2 & \qquad \longrightarrow \qquad \frac{2 S^2}{2} \geq \frac{S^{1}}{1} + \frac{S^{3}}{3} \\
i_3 & \qquad \longrightarrow \qquad 2S^3 + S^4 \geq S^1 + 2S^2 + S^5 \\
\textrm{cyclic inequality for odd $N$} & 
\qquad \longrightarrow \qquad N S^{(N+1)/2} \geq N S^{(N-1)/2} + S^N \\
i_5, i_6, i_7 & \qquad \longrightarrow \qquad \frac{2 S^3}{3} \geq \frac{S^{2}}{2} + \frac{S^{4}}{4} \\
i_8 & \qquad \longrightarrow \qquad 2 S^3 \geq 2S^1 + S^4
\end{align*}
The parent (detailed) inequalities are labeled following \cite{n5cuenca}. We list them for completeness in Appendix~\ref{n5cone}. Importantly, the symmetrizations of $i_1$ and $i_2$ and $i_{5,6,7}$ collectively imply the other symmetrized inequalities. 

\subsection{A strategy for proving the HCAE conjecture}

It is clear that inequality~$i_2$ and the trio~$i_{5,6,7}$ play a special role with regard to the HCAE conjecture. Once proven in their detailed form, they imply inequalities~(\ref{conjecture}) for $p = 2,3$ by symmetrization. We call such inequalities \emph{HCAE-realizing inequalities}. 

Up to now, the $p=2,3$ inequalities were the only proven inequalities in family~(\ref{conjecture}). The $p=4$ case is proven in this paper. Their $p\geq 5$ cousins remain a conjecture. In the three proven cases ($p=2,3,4$), the proof has proceeded by finding a detailed inequality and symmetrizing it. (For $p=2$ the symmetrization is trivial at $N=3$ because the parent inequality~(\ref{defmmi}) is already $S_4$-symmetric, but the comment is valid for $N>3$.) 

This suggests a potential strategy for proving the HCAE conjecture:

\paragraph{Strategy} Perhaps one can construct an HCAE-realizing inequality for every inequality~(\ref{conjecture}), using $N$ and $p$ as parameters. Thus far, efforts to find contraction maps for~(\ref{conjecture}) directly have failed. Perhaps the HCAE-realizing inequalities might have enough structure so that contraction maps for them could be constructed explicitly, even though the same task has remained elusive for their symmetrized descendants. \smallskip

The previous paragraph uses the word `perhaps' in two key places. One of our goals in writing this paper is to make the case that the strategy is nevertheless viable. We discuss this question in Section~\ref{sec:discussion}.  

\paragraph{A simplification} Inequalities~(\ref{conjecture}) depend on both $p$ and $N$. One technical simplification, which we emphasize from the start, is that it is enough to prove~(\ref{conjecture}) at every $p$ only for the smallest number of regions $N$, where the inequality is supposed to hold, that is for $N = 2p -1$.  Once proven there, the inequality follows for all $N > 2p - 1$ by symmetrizing over $S_{N+1}$. So, in fact, the strategy requires `only' to find a valid inequality with the structure
\begin{align}
(p-1) & (p+1) \times \big[\textrm{$p$-partite terms}\big] \label{structure} \\ 
& \geq \,\,\, \frac{p (p+1)}{2} \times \big[\textrm{$(p-1)$-partite terms}\big]
+ \frac{p(p-1)}{2} \times \big[\textrm{$(p+1)$-partite terms}\big] \nonumber
\end{align}
for every $p$. The inequality should involve exactly $N=2p-1$ regions. The structure~(\ref{structure}) can be modified by terms that symmetrize to zero. For instance, one can add an additional $p$-partite term on each side. As an example, this happens in inequality~$i_5$ of \cite{n5cuenca}, which realizes HCAE at $p=3$.

\subsection{In this paper:}
We take a first step in this program. 

In Section~\ref{sec:ineq}, we find explicitly an HCAE-realizing inequality for $p = 4$. This is the first known inequality for $N=7$ regions, other than the cyclic inequality that was discovered in \cite{hec}. It automatically proves the conjectured inequality~(\ref{conjecture}) for $p=4$. As such, it adds evidence in favor of the HCAE conjecture. Sections~\ref{sec:idea} and \ref{sec:route} explain the route by which we arrived at the inequality.

In Section~\ref{sec:properties}, we discuss properties of the new inequality. We emphasize properties which also apply to other holographic inequalities, in the hope that these may guide a search for higher-$p$ HCAE-realizing inequalities. We also establish that our inequality is the tightest possible, in the sense that one cannot subtract a term on the left or add a term on the right without invalidating it. 

Section~\ref{sec:discussion} is a discussion, which contains ideas for deploying our strategy at higher $p$. 

In the main text, we emphasize concepts and suppress details. Information, which is best presented as tables and graphs, is relegated to appendices. 

\section{New inequality}
\label{sec:ineq}

The inequality is:
\begin{align}
        S_{A B D E} + S_{A B D F}\, +\, & S_{A B E G} + S_{A D E F} + S_{A D E G} \nonumber \\
         +\, S_{A C D E} + S_{A C D F}\, +\, & S_{A C E G} + S_{B D E F} + S_{B D E G} \nonumber \\
         +\, S_{B C D E} + S_{B C D F}\, +\, & S_{B C E G} + S_{C D E F} + S_{C D E G} \label{ourineq} \\
        & \geq \nonumber \\
        S_{A B C} + S_{A D E} + S_{A D F} + S_{A E G} \, + \, & S_{B D E}
        + S_{B D F} + S_{B E G} \, + \,  S_{C D E} + S_{C D F} + S_{C E G} \nonumber \\
        +\, S_{A B D E F} + S_{A B D E G}  \, + \, & 
       S_{A C D E F} + S_{A C D E G} + 
        S_{B C D E F} + S_{B C D E G} \nonumber
\end{align}
It is an HCAE-realizing inequality for $p=4$. Referring to~(\ref{structure}), it has $3 \times 5 = 15 $ four-region terms on the left-hand-side, and $4 \times 5 / 2 = 10$ three-region terms and $4 \times 3 / 2 = 6$ five-region terms on the right-hand-side. As such, this inequality also proves inequality~(\ref{conjecture}) for $p=4$, which was conjectured in \cite{withsirui}.

We have confirmed the validity of~(\ref{ourineq}) using the software \cite{michael} on a standard Mac computer. We thank Michael Walter for making it available to us. 

\subsection{Motivating the search}
\label{sec:idea}
In the remainder of this section we explain how we arrived at inequality~(\ref{ourineq}). Readers who are uninterested in how the sausage gets made are advised to skip ahead to Section~\ref{sec:properties}.

In essence, we have been inspired by two facts: (i) that there are many more holographic entropy inequalities than there are entropies, which they constrain; and (ii) that known holographic entropy inequalities share many features in common. 

An example of fact (i) is the $N=5$ holographic entropy cone: it lives in a 31-dimensional space, but is bounded by 372 inequalities or facets \cite{n5cuenca}. The gap between the number of inequalities and the dimension of entropy space is expected to be even larger at higher $N$. This implies that the positive quantities defined by the inequalities necessarily share many linear dependencies. If so, at least in some lucky cases it should be true that an $N=7$ inequality will be expressible as a linear combination of lower-$N$ inequalities.\footnote{Technically, we are speaking here about linear combinations not of inequalities, but of positive quantities defined by the inequalities. The distinction is that the latter allows negative coefficients. Examples of the type of linear combinations we are after include inequalities~(\ref{i7rewriting}), (\ref{ourconstr}), and (\ref{n5cyclicrewrite}).} We are interested in finding at least one such instance, which is HCAE-realizing. 

Fact~(ii)---that maximally tight holographic entropy inequalities share many features---makes it logical to harness known inequalities when searching for new ones. Features shared by all holographic entropy inequalities include superbalance \cite{superbalance} (see beginning of Section~\ref{sec:properties}) and positivity in the K-basis \cite{kbasis} (see Appendix~\ref{sec:kbasis}). We list a few further properties in Section~\ref{sec:properties}. If we are to build a new inequality that satisfies certain conditions, it is efficient to use building blocks, which already satisfy the same conditions.

\paragraph{Prior example}
To illustrate the idea, consider inequality~$i_7$ from \cite{n5cuenca}; see Appendix~\ref{n5cone} for its explicit form. It can be rewritten in the following way:
\begin{equation}
\!\!\! I(AD:B:E) + I(A:BE:D) + I(AD:BE:C) \geq I(A:C:E) + I(ACE:B:D)~~
\label{i7rewriting}
\end{equation}
Here $I(X:Y:Z)$ is the positive quantity defined by the monogamy inequality~(\ref{defmmi}):
\begin{equation}
I(X:Y:Z) \equiv S_{XY} + S_{YZ} + S_{ZX} - S_X - S_Y - S_Z - S_{XYZ}
\label{defi3}
\end{equation}
This quantity is often called $-I_3(X:Y:Z)$. For the purposes of this paper, we define $I(X:Y:Z)$ without the minus sign so it is easier to keep track of positive quantities. In effect, we see that inequality~$i_7$ compares different lifts of the monogamy of mutual information to the context of $N=5$ regions.

Inequality~$i_7$ is an HCAE-realizing inequality for $p=3$. The monogamy of mutual information---the inequality used in rewriting~(\ref{i7rewriting})---is the HCAE-realizing inequality for $p=2$. We surmised---correctly---that an HCAE-realizing inequality for $p=4$ could be constructed in a similar fashion by combining different lifts of $p=3$ HCAE-realizing inequalities.

\subsection{Working toward the inequality}
\label{sec:route}
We took inequality~$i_5$ as the starting point in our guesswork. The other HCAE-realizing inequalities for $p=3$ are $i_6$ and $i_7$. However, it turns out that using only $i_5$ as a building block is sufficient to construct a $p=4$ HCAE-realizing inequality. 

The explicit form of $i_5$ is:
\begin{align}
S_{ABD} + S_{ACD} + S_{BCD} + S_{ABE} \, + & \, S_{ACE} + S_{BCE} + S_{ADE} + S_{BDE} + S_{CDE}  \nonumber \\
& \geq \label{i5} \\
S_{AD} + S_{AE} + S_{BD} + S_{BE} + S_{CD} \, + \, & S_{CE} 
+ S_{ABC} + S_{ABDE} + S_{ACDE}  + S_{BCDE} \nonumber
\end{align}
In the context of $N=5$ named regions, this inequality has symmetry group $S_3 \times S_3 \times \mathbb{Z}_2$. The first $S_3$ is explicit in (\ref{i5}): it comprises permutations of regions $\{A,B,C\}$. The second $S_3$ is obscured by the presentation in (\ref{i5}), where the purifying region $O$ is never written down, but it is easy to confirm. It comprises permutations of regions $\{D,E,O\}$. The $\mathbb{Z}_2$ swaps the two triples of regions.

\paragraph{Classifying lifts of $i_5$ by $S_8$-invariants}
The initial step in the construction is to identify $N=7$ lifts of $i_5$, which are distinct in terms of how many $q$-partite regions they contain. The composition of the inequality in terms of $q$-partite components is known from (\ref{structure}). Therefore, this analysis informs us how many distinct lifts should be combined, and in what proportion.

Here we are lifting to $N=7$ named regions an inequality, which is initially characterized by two symmetry-equivalent triples of regions: $\{A,B,C\}$ and $\{D,E,O\}$. We are needing to adjoin two additional regions $F,G$ in some way. Treating all regions (including the purifier $O$) on an equal footing, we have three $S_8$-inequivalent ways of doing so:
\begin{enumerate}
\item Adjoining $F,G$ to a single region. Example: $O \to OFG$.
\item Adjoining $F$ to one region and $G$ to another, both from the same triple. Example: $D \to DF$ and $O \to OG$. 
\item Adjoining $F$ to one region and $G$ to another, taken from distinct triples. Example: $A \to AF$ and $O \to OG$.
\end{enumerate} 
These three lifts symmetrize in the $N=7$ context (where $S^3 = S^5$) to the following combinations:
\begin{align}
8 S^3 & \geq 6 S^2 + 3 S^4 \\
6 S^4 & \geq 3 S^2 + 4 S^3 \\
3 S^4 & \geq 4 S^2
\end{align}
We are looking for an inequality that symmetrizes to:
\begin{equation}
\!\!\!\!\!\!\!\!\!\!\!
\textrm{ineq.~(\ref{structure}) at $p=4$} \qquad \longrightarrow \qquad 15 S^4 \geq 16 S^3
\end{equation}
If the three distinct lifts of $i_5$ are to assemble it, they must appear with coefficients that solve the following system of equations:
\begin{align}
-6a -3 b - 4c & = 0 \nonumber \\
8a -4b \phantom{\phantom{\scalebox{.0001}{|}} + 3c} & = -16 \\
-3a + 6b + 3c & = + 15 \nonumber
\end{align}
There are infinitely many solutions:
\begin{equation}
(a,b,c) = (-1, 2, 0) + \textrm{(any coefficient)} \times (1,2,-3).
\label{linearsol}
\end{equation}
The homogeneous term $\Delta (a,b,c) = (+1,+2,-3)$ has a vanishing symmetrization at $N=7$ regions. It cannot be fixed by inspecting permutation invariants. 

\paragraph{Guesswork} For an initial attempt, we took the coefficients $(a,b,c) = (-1,2,0)$. There is no compelling reason to do so, but for human researchers it is the simplest and most intuitive option. It is easier to work with two non-vanishing coefficients instead of three. But setting $a=0$ or $b=0$ would mean that the triple-region content of the intended inequality would be taken entirely from one lift of $i_5$---an option we considered unlikely. 

We proceeded to inspect combinations of the form $+2 \times \textrm{(lift~2.)} \geq 1 \times \textrm{(lift~1.)}$. We looked for combinations of this type, in which double-region entropies cancel out identically. The remaining choice at this stage was to select the exact lifts. (The list on the previous page classifies $S_8$-inequivalent lifts, but now we are looking for the exact inequality, so the particular representatives of the $S_8$-equivalent classes matter.) 

To explain the choice, we need a notation for the positive quantity defined in inequality $i_5$. (This definition mirrors equation~(\ref{defi3}) from the `Prior Example.') We let:
\begin{align}
& I(X:Y:Z;~V:W) \label{defirrrrr} \\
= \, & \phantom{\Big(} S_{XYV} + S_{XZV} + S_{YZV} + S_{XYW} + S_{XZW} + S_{YZW} + S_{XVW} + S_{YVW} + S_{ZVW} \nonumber \\
- \, & \Big(  S_{XV} + S_{XW} + S_{YV} + S_{YW} + S_{ZV} + S_{ZW} 
+ S_{XYZ} + S_{XYVW} + S_{XZVW} + S_{YZVW}  \Big) \nonumber
\end{align} 
In this notation, lift~2. can take the form $I(RR:RR:R;~R:R)$ or $I(R:R:R;~RR:R)$ or $I(R:R:R;~RR:RR)$, depending on whether we adjoin regions $F,G$ to regions from triple $\{A,B,C\}$ or $\{D,E,O\}$, and in the latter case whether or not we adjoin to the purifier. Lift~1., on the other hand, can take the form $I(R:R:R;~R:R)$ or $I(RRR:R:R;~R:R)$ or $I(R:R:R;~RRR:R)$, depending on whether we adjoin regions $FG$ to one of $\{A, B, C\}$, to one of $\{D, E\}$, or to the purifier $O$. The options for each lift map to one another under the permutation group $S_8$, but become distinct when we wish to distinguish the purifier $O$ from the seven named regions. 

If we assume that the inequality really involves $+2$ copies of lift~2. and $-1$ copy of lift~1. (as opposed to, say, $+4$ and $-2$ copies) then we can set the option for lift~1. at will. We chose the option $I(R:R:R;~R:R)$, which appeared the simplest to the human eye. Without loss of generality, we are therefore looking for:
\begin{equation}
2 \times \textrm{(lift~2.)} \geq I(A:B:C;~D:E)
\end{equation}
To select the right choice of lifts for the right hand side, we now demand that two-region entropies and six-region entropies cancel out from the inequality identically. (We need not consider single- or seven-region entropies because they do not appear in lifts~2.) This demand, too, is a guess: it is possible that two-region entropies can appear on both sides of the inequality, but cancel out from the symmetrization. One rationale for attempting this was that inequalities $i_5, i_6, i_7$ do not feature any single- or five-region terms, which play analogous roles at $p=3$. 

This demand turns out to be a strong contraint. It has an essentially unique\footnote{There is one other solution: $I(A:B:C;~DF:E) + I(A:B:C;~D:EF) \stackrel{?}{\geq} I(A:B:C;~D:E)$. However, that guess cannot be correct because it does not involve $G$ at all and we are looking for an inequality on $N=7$ regions. Explicitly, this guess is disproved by the six-party perfect tensor state on $A,B,C,D,E,F$. Perfect tensors are defined in equation~(\ref{defpt}) below. \label{wrongguess}} solution:
\begin{equation}
I(A:B:C;~DF:E) + I(A:B:C;~D:EG) \geq I(A:B:C;~D:E)
\label{ourconstr}
\end{equation}
Substituting definition~(\ref{defirrrrr}) gives inequality~(\ref{ourineq}). We then verified that~(\ref{ourineq}) is valid using software \cite{michael}. 

\section{Properties of the new inequality}
\label{sec:properties}
All valid holographic inequalities are known to enjoy a property called superbalance \cite{superbalance}. It says that every individual region appears an equal number of times on both sides of the inequality. The `super-' part in superbalance posits that this includes the purifier $O$. As a consistency check, we can easily confirm that inequality~(\ref{ourineq}) satisfies this condition. In fact, superbalance of (\ref{ourineq}) follows directly from the rewriting~(\ref{ourconstr}) because the $i_5$-ingredients are each superbalanced.  

We now discuss other features of inequality~(\ref{ourineq}), emphasizing those which are shared with previously known holographic inequalities. We hope that these features will be helpful in constructing further inequalities and/or in proving the HCAE conjecture. 

\subsection{Reductive property---applies to all inequalities}
\label{sec:reduction}

Whenever you set one region to be empty (or unentangled), every holographic inequality for $N$ regions reduces to another valid inequality on fewer regions. Heuristically, this property sets a counterpoint to the method by which we guessed inequality~(\ref{ourineq}): reductions go from higher to lower $N$ whereas we considered lifts from lower to higher $N$. This is why we inspect them in this paper.  

Some remarks:
\begin{itemize}
\item The reductive property includes setting the purifier $O$ to be unentangled, which means that the $N$ named regions land in a pure state by themselves. 
\item In all hitherto known cases, reducing an $N$-region inequality produces a $(N-2)$-region inequality, possibly applied to one composite region such as $AB$. 
\item Inequalities $i_6$ and $i_8$ from the $N=5$ holographic entropy cone reduce to convex combinations of $N=3$ inequalities (two instances of monogamy).
\item We allow $0 = 0$ as a possible reduction. 
\end{itemize}
\smallskip

\paragraph{Example~1: Monogamy of mutual information} When we set $S_A = 0$, inequality~(\ref{defmmi}) reduces to $0 = 0$. The same is true for the other reductions because monogamy of mutual information is $S_4$-symmetric. 

\paragraph{Example~2: Inequality~$i_5$} The inequality is given in (\ref{i5}) and in Appendix~\ref{n5cone}. Setting $S_A = 0$ reduces $i_5$ to an instance of the monogamy of mutual information:
\begin{equation}
S_A = 0 \quad \Longrightarrow \quad I(BC:D:E) \geq 0
\label{i5reduce}
\end{equation}
Here we used the notation of equation~(\ref{defi3}). All other reductions are also instances of monogamy because the symmetry of $i_5$ allows us to swap $A$ with any other region; viz. the discussion below~(\ref{i5}). 

\paragraph{Example~3: The new inequality} Let us verify its reductive property using the rewriting~(\ref{ourconstr}). There are two symmetry-inequivalent reductions to consider: $S_D = 0$ and $S_A = 0$. (See Section~\ref{sec:symmetry} below for the symmetry analysis.) We use the reduction of $i_5$, which was displayed in (\ref{i5reduce}), in all steps. 

The reduction $S_D = 0$ directly produces $I(A:B:C;~E:F) \geq 0$, an instance of $i_5$.

The reduction $S_A = 0$ gives:
\begin{equation}
I(BC:DF:E) + I(BC:D:EG) \geq I(BC:D:E) \label{n5cyclicrewrite}
\end{equation}
In Appendix~\ref{allreductions} we verify that (\ref{n5cyclicrewrite}) represents the cyclic inequality on the ordered set of regions $D$-$F$-$O$-$G$-$E$, with the composite region $BC$ treated as the purifier. Incidentally, rewriting~(\ref{n5cyclicrewrite}) shows that the $N=5$ cyclic inequality could also be found using the method, which we employed in Section~\ref{sec:route}. This adds a little circumstantial motivation for our search strategy. 
\smallskip

Appendix~\ref{allreductions} lists the reductions of the other known holographic entropy inequalities: the cyclic inequalities, $i_6$, $i_7$, and $i_8$. We tabulate them in the hope that they will be useful in implementing our strategy in a more exhaustive search for holographic inequalities in the future.  

\subsection{Hyperbalance---applies to HCAE-realizing inequalities}
\label{sec:hyper}

We now know five HCAE-realizing inequalities: $i_2$ for $p=2$, $i_5, i_6, i_7$ for $p=3$, and the new inequality~(\ref{ourineq}) for $p=4$. All have the composition~(\ref{structure}). 

Each of them displays an additional interesting feature: every region enters the $(p-1)$-partite terms and the $(p+1)$-partite terms on the right hand side the same number of times. For example, in the monogamy of mutual information (which realizes HCAE for $p=2$) we have single-region terms $S_A + S_B + S_C$ and one triple-region term $S_{ABC}$. We propose to call this feature `hyperbalance,' in contrast to the previously observed and proven superbalance \cite{superbalance}. It is important to stress that hyperbalance relates terms on the same side of an inequality whereas superbalance relates terms on different sides of an inequality. 

We will say that an inequality is fully hyperbalanced if the said property holds with respect to all the regions, including the purifier. To verify hyperbalance with respect to $O$, we rewrite the inequality in such a way that a previously named region assumes the role of the purifier. For example, to make $A$ into the purifier, we flip every term that features $A$ for its complement, e.g. $S_{ABC} \to S_{O}$ and $S_{A} \to S_{BCO}$ in the $N=3$ context. Doing so for our new inequality gives:
\begin{align}
        S_{CFGO} + S_{CEGO}\, +\, & S_{CDFO} + S_{BCGO} + S_{BCFO} \nonumber \\
         +\, S_{BFGO} + S_{BEGO}\, +\, & S_{BDFO} + S_{BDEF} + S_{BDEG} \nonumber \\
         +\, S_{BCDE} + S_{BCDF}\, +\, & S_{BCEG} + S_{CDEF} + S_{CDEG} \label{apure} \\
        & \geq \nonumber \\
        S_{BDE} + S_{BDF} + S_{BEG} + S_{BFO} \, + \, & S_{BGO} +
        S_{C D E} + S_{C D F} + S_{C E G} + S_{CFO} + S_{CGO} \nonumber \\
        +\, S_{B C D E F} + S_{B C D E G} \, + \, & S_{BCDFO}  
        + S_{BCEGO} + S_{BCFGO} + S_{DEFGO} \nonumber
\end{align}
Again, each region appears the same number of times in the three-partite and five-partite terms. The purifier flip $A \leftrightarrow O$ is the only non-trivial one, up to symmetry (see below). 

We have also verified hyperbalance for the other HCAE-realizing inequalities $i_5, i_6, i_7$. 

Below we explain hyperbalance in terms of extreme vectors of the holographic entropy cone, which saturate inequalities~(\ref{structure}). To do this, we need to visualize entropy vectors in terms of graph models. A way to do so was explained in Reference~\cite{hec}.

\paragraph{Graph models for entropy vectors} Draw a graph with weighted edges. Label a subset of its vertices $A, B, C \ldots O$; these vertices will be collectively called \emph{external}. For every $X \subset \{A, B, C, \ldots O\}$, identify the minimum total weight of edges, which must be cut if $X$ is to be separated from the remaining external vertices. Let this number---the total weight of the minimal cut---become the entropy $S_X$. This is how a weighted graph defines an allowed set of entanglement entropies. After symmetrization, it also defines a set of permutation-invariants $S^p$. Reference~\cite{hec} proved that any set of entropies constructed in this way can be realized as areas of Ryu-Takayanagi surfaces in a classical geometry. 

\begin{figure}[tbp]
\centering 
\includegraphics[width=.6\textwidth]{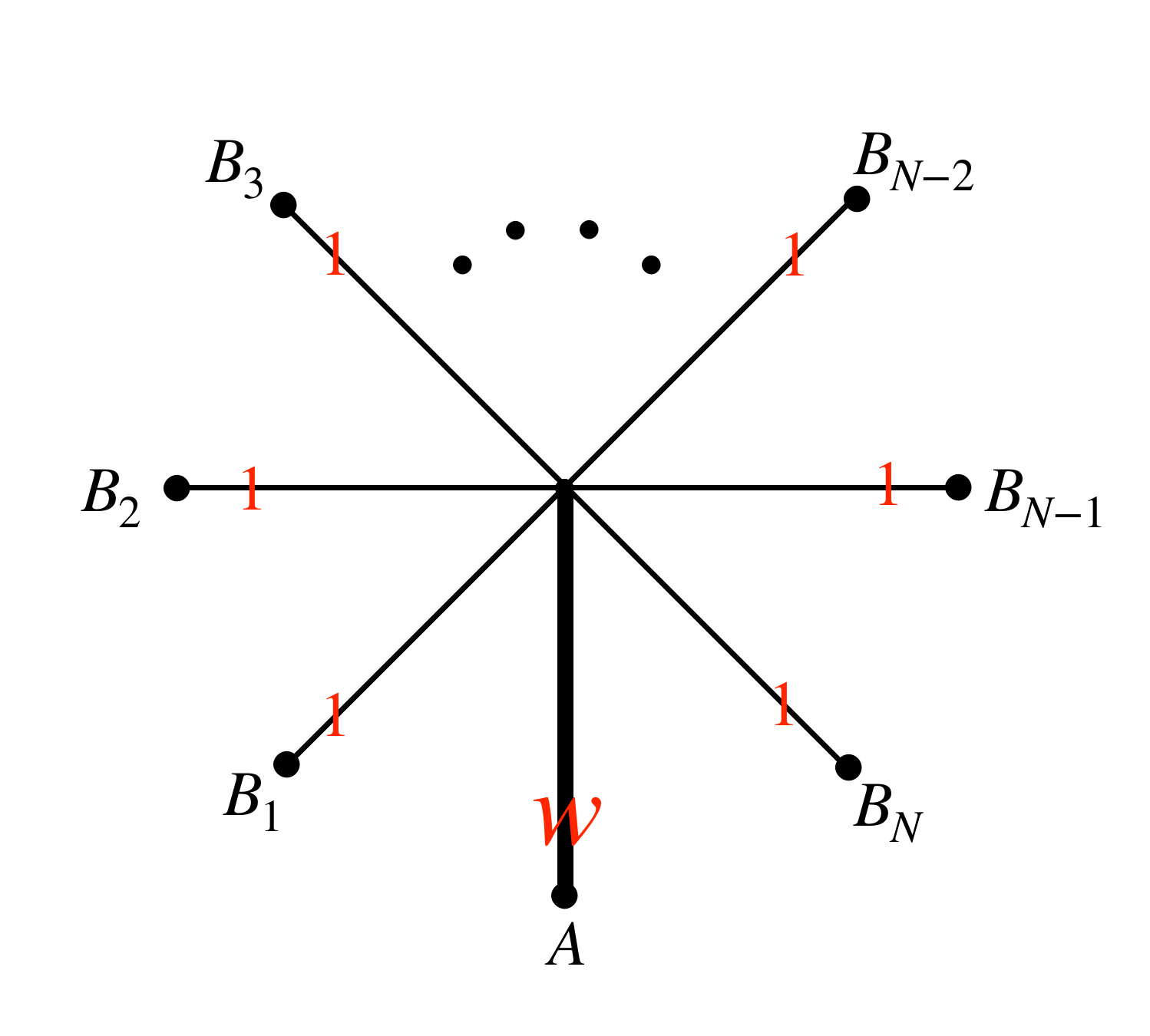}
\hfill
\caption{The graph model of extreme rays of the holographic entropy cone, whose symmetrized entropies $S^p$ are extreme in the holographic cone of average entropies (HCAE). For $N$ odd (as is assumed in this paper), the parameter $w$ takes odd values between 1 and $N$. The graph was drawn following \cite{withsirui}.}
\label{fig:flowers}
\end{figure}

\paragraph{Explanation of hyperbalance} Reference~\cite{withsirui} identified graph models whose entropic invariants $S^p$ form extreme rays in the HCAE. Their extremality means that they simultaneously saturate as many inequalities~(\ref{conjecture}) as possible. They are shown in Figure~\ref{fig:flowers}. An interesting aside is that the same graph models characterize the unitary evaporation of black holes; see \cite{withsirui, octopi}.

Because we are interested in inequalities~(\ref{structure}), we set $N = 2p-1$. The relevant graphs are star graphs with $2p$ legs, with all but one leg of weight 1. In the graphs, which define extreme rays in the HCAE, the special weight $w$ takes odd values between 1 and $N$. Note that every $w\geq 3$ defines an $(N+1)$-tuple of extremal vectors because the differently weighted edge can be attached to any region. The $w=1$ option is a singlet under $S_{N+1}$; it defines a so-called perfect tensor state.

The perfect tensor on $2p$ constituents does not saturate inequality~(\ref{structure}). It is easy to verify this using its entanglement entropies:
\begin{equation}
S_{\textrm{(any~$p\pm 1$~regions})} = p-1
\qquad {\rm and} \qquad
S_{\textrm{(any~$p$~regions})} = p
\label{defpt}
\end{equation}
But the other configurations of Figure~\ref{fig:flowers} saturate (\ref{structure}) if it is hyperbalanced. To confirm this, suppose that the special vertex attached to the $w$-weighted edge is $A$, as displayed in the figure. The total number of legs is $N+1 = 2p$. In this case, the entanglement entropies become:
\begin{align}
S_{\textrm{($p-1$~regions~excluding~$A$})} & = p-1
& \qquad
S_{\textrm{($p-1$~regions~including~$A$})} & = p+1
\nonumber \\
S_{\textrm{($p$~regions)}} & = p
& \qquad
&
\\
S_{\textrm{($p+1$~regions~excluding~$A$})} & = p+1
& \qquad
S_{\textrm{($p+1$~regions~including~$A$})} & = p-1
\nonumber
\end{align}
Substitute these expressions in inequality~(\ref{structure}). After all the cancelations, we obtain:
\begin{equation}
0 \geq 2\, \#\{ (p-1)\textrm{-partite terms containing $A$}\} - 2\, \#\{ (p+1)\textrm{-partite terms containing $A$}\}
\end{equation}
Therefore, the configuration in Figure~\ref{fig:flowers} (with $w \geq 3$) saturates inequality~(\ref{structure}) if the latter is hyperbalanced with respect to $A$. 

\subsection{Symmetry}
\label{sec:symmetry}
The symmetry group of our new inequality is $S_3 \times D_5$. In the presentation~(\ref{ourineq}), the $S_3$ permutes regions $\{A,B,C\}$. The dihedral symmetry $D_5$, which reshuffles the remaining regions, can be visualized by placing the regions on vertices of a regular pentagon in the cyclic order  $E$-$F$-$G$-$D$-$O$. 

The $S_3$ acting on $\{A,B,C\}$ is manifest in (\ref{ourineq}). The $D_5$ component is harder to see in this presentation, but it becomes manifest in the K-basis \cite{kbasis}. We briefly review the K-basis, and rewrite our inequality using it, in Appendix~\ref{sec:kbasis}.

The symmetry group is of order 60. This means that our inequality defines $8! / 60 = 672$ distinct facets of the $N=7$ holographic entropy cone. This is assuming that the inequality cannot be improved, so it indeed marks a facet of the cone. We verify this presently. 

\subsection{Our inequality cannot be improved}
\label{sec:tight}

Our inequality is maximally tight. This means that one cannot subtract a term from the `greater than' side of the inequality, or add a term to the `less than' side of the inequality, without violating it. The locus where a maximally tight inequality is saturated forms a facet (bounding codimension-one hyperplane) of the holographic entropy cone. Inequality~(\ref{ourineq}) is a facet of the $N=7$ cone.

To show that the inequality cannot be improved, we must demonstrate that it can be saturated in 126 linearly independent ways. The 126 follows because the $N=7$ entropy space is $(2^7 - 1 = 127)$-dimensional, and the saturation locus is supposed to be codimension-one. 

\begin{figure}[t]
\centering 
\includegraphics[width=.99\textwidth]{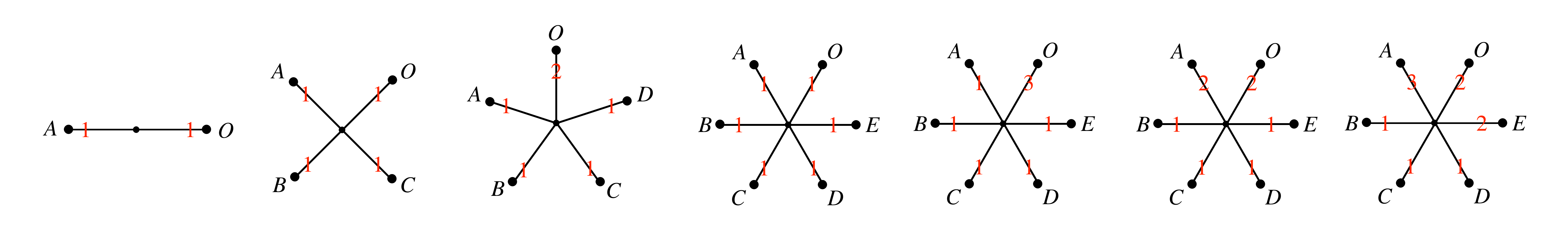}
\hfill
\caption{Graphs models of extreme vectors of the $N=5$ holographic entropy cone, which are used in Section~\ref{sec:tight}. The graphs were drawn following \cite{n5cuenca}.}
\label{fig:stars}
\end{figure}

We took all the extreme vectors of the $N=5$ holographic entropy cone \cite{n5cuenca} whose graph models (see Section~\ref{sec:hyper}) are star graphs, and lifted them to the $N=7$ context. Those graphs are shown in Figure~\ref{fig:stars}. After accounting for permutations, the lifts give rise to 2674 configurations. Of those, 903 saturate our inequality. We found that 125 of them are linearly independent. In addition, we knew from `Explanation of hyperbalance' that the graphs from Figure~\ref{fig:flowers} at $N=7$ also saturate it. We confirmed that each graph from Figure~\ref{fig:flowers} with $w=3$ or $w=5$ is linearly independent of the saturating configurations lifted from the $N=5$ cone. The calculations were conducted using Mathematica. 

\section{Discussion}
\label{sec:discussion}

In Section~\ref{sec:reduction}, we considered how basic holographic inequalities reduce when one region is removed or unentangled. They produce lower-$N$ holographic inequalities. (In the cases $i_6$ and $i_8$, they produce convex combinations of basic inequalities.) The method by which we constructed inequality~(\ref{ourineq}) was, in essence, the undoing of one such reduction.\footnote{Following common chemical nomenclature, we propose to call this method `oxidation.'} We assumed that at least one inequality of the form~(\ref{structure}) would reduce to $i_5$ after setting $S_F = 0$ and $S_G = 0$. We then considered various ways of adjoining $F$ and $G$ to $I(A:B:C;~D:E) \geq 0$ and immediately came up with (\ref{ourconstr}). 

Inequality~(\ref{ourineq}) was only the second option we checked. (We first inspected the option in footnote~\ref{wrongguess}, which in retrospect could not have worked.) The fact that we generated a valid inequality already on the second trial---and the first one that did not violate an obvious requirement---suggests that the `oxidation' technique can be employed more broadly. We anticipate that many new holographic inequalities can be generated by the same procedure.

We offer a few comments about how a more complete search can and should differ from the logic and guesswork outlined in Section~\ref{sec:route}:
\begin{itemize}
\item We used $i_5$ as the only building block. In a more exhaustive search, we would need to consider different ones. Inequalities $i_6$ and $i_8$ illustrate this possibility because [some of] their reductions involve linear combinations of basic lower-$N$ inequalities.
\item For inequality~(\ref{ourineq}), the guess was tightly constrained by the anticipated structure~(\ref{structure}). Searching for inequalities, which are not HCAE-realizing, will not have this benefit.
\item However, our search for~(\ref{ourineq}) only considered adjoining $F$ and $G$ to a fixed starting point $I(A:B:C;~D:E) \geq 0$. We inspected how the outcome reduces when other regions $R$ decouple ($S_R = 0$) only after we found inequality~(\ref{ourineq}). It seems promising to construct inequalities by assuming that they have fixed reductions on several regions. For example, if we had assumed that the new inequality reduces to $I(A:B:C;~E:F) \geq 0$ when $S_D = 0$ \emph{and} that it reduces to the cyclic inequality on $D$-$F$-$G$-$O$-$E$ when $S_A = 0$, we would have found it even without assuming (\ref{structure}). 
\end{itemize}
We expect that our method can generate other facets of the $N=6$ and $N=7$ holographic entropy cones. Beyond that, the program will run into the difficulty of verifying candidate inequalities. Barring some game-changing insight, verifying inequalities by brute-searching for contractions will not succeed far beyond $N=7$. 

HCAE-realizing inequalities appear to be one promising exception, where this idea might succeed for arbitrary $N$. This is because their structure is so tightly constrained: they must be of the form~(\ref{structure}) and they must be hyperbalanced (Section~\ref{sec:hyper}). These facts may contain enough hints to formulate a regular family of HCAE-realizing inequalities for every $N$. The cyclic family of \cite{hec} is one prototype of a regular, infinite family of inequalities. 

Concretely, the following inductive argument seems feasible. Perhaps an $(N+2)$-region HCAE-realizing inequality can be constructed by lifting the $N$-region HCAE-realizing inequality (assumed known as the inductive hypothesis) and/or the $N$-region cyclic inequality. If such a setup could be correctly invented, presumably the relevant contraction maps could also be constructed analytically. Here, again, the contraction maps for the cyclic inequalities provide an inspirational prototype. We intend to explore this possibility in future work.  

\paragraph{Relation to prior work} The recent reference \cite{sacone} formulated a program for finding holographic entropy inequalities, which uses the subadditivity of entanglement entropy to bootstrap one's way from lower-$N$ to higher-$N$ inequalities. That program has some---thus far unexplored---relation with the HCAE conjecture because subadditivity is a measure of bipartite correlations. On the other hand, \cite{withsirui} established that the EPR-like behavior of $(p' \geq p)$-partite entanglement entropies characterizes the most efficient holographic erasure-correcting schemes, which correct every $p$-region erasure.

It would be interesting to study in detail how the work of \cite{sacone} intersects the HCAE conjecture and, by extension, the content of this paper. Another way to organize and predict the structure of holographic entropy inequalities appeared in \cite{arrangement}. We hope to elucidate its relation to the present work in the future.  

\section*{Acknowledgments} 

We thank Michael Walter for making his code \cite{michael} available and explaining it to us. We also thank Sergio Hern\'andez Cuenca, Massimiliano Rota, Sirui Shuai and Daiming Zhang for useful conversations, and Sirui Shuai for sharing handy Mathematica code. 

\appendix
\section{Previously known holographic inequalities}
\label{n5cone}

The inequalities are listed in \cite{n5cuenca} with labels $i_1$ through $i_8$. We stick to the same labeling below.
\begin{itemize}
\item Inequality~$i_1$ is subadditivity:
\begin{equation}
S_A + S_B \geq S_{AB} \tag{$i_1$}
\end{equation}
\item Inequality~$i_2$ is the monogamy of mutual information:
\begin{equation}
S_{AB} + S_{BC} + S_{AC} \geq S_A + S_B + S_C + S_{ABC} \tag{$i_2$}
\end{equation}
\item Inequality~$i_3$ is the monogamy of mutual information, with two extra regions $D$ and $E$ adjoined to two distinct regions. Here we adjoin $B \to BD$ and $C \to CE$:
\begin{equation}
S_{ABD} + S_{BCDE} + S_{ACE} \geq S_A + S_{BD} + S_{CE} + S_{ABCDE} \tag{$i_3$}
\end{equation}
\item Inequality~$i_4$ is part of an infinite family of cyclic inequalities, which were first presented in \cite{hec}:
\begin{align}
S_{A_1 A_2 \ldots A_p} + S_{A_2 A_3 \ldots A_{p+1}} & \, + \, \ldots + S_{A_{2p-1} A_1 \ldots A_{p-1}}
\nonumber \\
& \geq \tag{$i_4$} \label{gencyclic} \\
S_{A_1 A_2 \ldots A_{p-1}} + S_{A_2 A_3 \ldots A_{p}} \, + \, & \ldots + S_{A_{2p-1} A_1 \ldots A_{p-2}}
+ S_{A_1 A_2 \ldots A_{2p-2} A_{2p-1}}
\nonumber
\end{align}
The regions $A_i$, which are indexed with $1 \leq i \leq 2p-1$, are cyclically ordered, so $A_1 \equiv A_{2p}$. The ellipses in subscripts denote unions of consecutively indexed regions. We refer to (\ref{gencyclic}) as the $N=2p-1$ cyclic inequality. Note that the $N=3$ instance is the monogamy of mutual information $i_2$. 
\item Inequality~$i_5$ is:
\begin{align}
S_{ABD} + S_{ACD} + S_{BCD} + S_{ABE} \, + & \, S_{ACE} + S_{BCE} + S_{ADE} + S_{BDE} + S_{CDE}  \nonumber \\
& \geq \tag{$i_5$} \\
S_{AD} + S_{AE} + S_{BD} + S_{BE} + S_{CD} \, + \, & S_{CE} 
+ S_{ABC} + S_{ABDE} + S_{ACDE}  + S_{BCDE} \nonumber
\end{align}
\item Inequality~$i_6$ is:
\begin{align}
\!\!\!\!\!
3S_{ABC} + 3S_{ABD} + S_{ABE} + S_{ACD} + 3 S_{ACE} \,+\, & 
S_{ADE} + S_{BCD} + S_{BCE} + S_{BDE} + S_{CDE} \nonumber \\
& \geq \tag{$i_6$} \\
2S_{AB}+ 2S_{AC} + S_{AD}+ S_{AE} \,+\, & S_{BC}+ 2S_{BD}+ 2 S_{CE}+ S_{DE} \nonumber \\
+\, 2 S_{ABCD} + 2S_{ABCE} \,+\, & S_{ABDE} + S_{ACDE} \nonumber
\end{align}
This inequality has an $S_3 \times \mathbb{Z}_2$ symmetry. The $S_3$ permutes ordered pairs $\{(A,O)$, $(E,B)$, $(D,C)\}$ into one another whereas the $\mathbb{Z}_2$ simultaneously switches the ordering in those pairs.
\item Inequality~$i_7$ is:
\begin{align}
2S_{ACE} + S_{ABD} + S_{ABE} \, + & \, S_{ADE} + S_{ACD} + S_{BCE} + S_{BDE} \nonumber \\
& \geq \tag{$i_7$} \\
S_{AC} + S_{AD} + S_{AE} + S_{BD} \, + \, & S_{BE} + S_{CE} 
+ S_{ABCE} + S_{ABDE}  + S_{ACDE} \nonumber
\end{align}
This form differs from the one presented in \cite{n5cuenca} by the exchange $B \leftrightarrow E$. The advantage of this labeling is that now a $\mathbb{Z}_6$ symmetry permutes $(ABCDEO)$ in this order. We used this form when rewriting $i_7$ in inequality~(\ref{i7rewriting}). 
\item Inequality~$i_8$ is:
\begin{align}
S_{AD} + S_{BC} + S_{ABE} + S_{ACE} & + S_{ADE} + S_{BDE} + S_{CDE} \nonumber \\
\geq & \tag{$i_8$} \\ 
S_{A} + S_{B} + S_{C} + S_{D} + S_{AE} + S_{DE} & + S_{BCE} + S_{ABDE} + S_{ACDE} \nonumber
\end{align}
This inequality has $\mathbb{Z}_2$ symmetries, which switch $A \leftrightarrow D$ and $B \leftrightarrow C$, and another one that switches the ordered triples $(A,D,E)$ and $(B,C,O)$ with one another. 
\end{itemize}

\section{Reductions of known inequalities}
\label{allreductions}

The reductions of $i_1$ and $i_2$ are $0=0$. Inequality~$i_3$ is already an instance of $i_2$ for a composite region. We showed in the main text that $i_5$ reduces to the monogamy of mutual information. Here we inspect the other known holographic entropy inequalities. We use the notation $I(X:Y:Z)$ defined in equation~(\ref{defi3}) throughout. 

\paragraph{The cyclic inequalities} 
When we set the purifier to be unentangled, inequality~($i_4$) reduces to $0 = 0$. In that circumstance terms on both sides become equal pairwise because they describe complements in a pure state on $\cup_{i=1}^{2p-1} A_i$. 

Up to cyclic symmetry, the only other reduction is to set one of the $2p-1$ named regions unentangled. We claim that this reduces ($i_4$) at parameter $p$ to the same inequality at parameter $p-1$. 

Without loss of generality, let the unentangled system be $A_p$. Then terms $S_{A_1 A_2 \ldots A_p}$ and $S_{A_p A_{p+1} \ldots A_{2p-1}}$ on the left hand side cancel out with $S_{A_1 A_2 \ldots A_{p-1}}$ and $S_{A_{p+1} A_{p+2} \ldots A_{2p-1}}$ on the right. After the reduction, the left hand side contains $(p-1) \times$ [$p$-partite terms] and $(p-2) \times $ [$(p-1)$-partite terms]. The former all contain $A_1 A_{2p-1}$ in combination while the latter do not contain them at all. The right hand side, in addition to $S_{A_1 \ldots A_{p-1} A_{p+1} \ldots A_{2p-1}}$, now contains $(p-2) \times$ [$(p-1)$-partite terms] and $p \times$ [$(p-2)$-partite terms]. The former all contain $A_{1} A_{2p-1}$ in combination whereas the latter do not contain them at all. 

We now set $A_1 A_{2p-1} \equiv B$. Written in terms of $B$, all terms on the left become $(p-1)$-partite and all terms on the right (except $S_{B A_2 \ldots A_{p-1} A_{p+1} \ldots A_{2p-2}}$) become $(p-2)$-partite. If we place $B$ between $A_{2p-2}$ and $A_{2}$ in a cyclic ordering then all terms describe consecutive unions. These properties uniquely identify the cyclic inequality on $N = 2(p-1) - 1$ regions.

\paragraph{Inequality~$i_6$} Because the symmetry of $i_6$ can map any region into any other, it is enough to set $S_A = 0$. This reduces $i_6$ to a sum of two monogamies:
\begin{equation}
I(B:C:D) + I(B:C:E) \geq 0
\end{equation}

\paragraph{Inequality~$i_7$} Setting $S_A=0$ in inequality~$i_7$ gives $I(C:D:E) \geq 0$. The other reductions are also instances of monogamy because $i_7$ has a cyclic symmetry, which rotates all regions into one another, including the purifier. 

It is interesting to examine this reduction using rewriting~(\ref{i7rewriting}), which we reproduce here for the reader's convenience: 
\begin{equation*}
I(AD:B:E) + I(A:BE:D) + I(AD:BE:C) \geq I(A:C:E) + I(ACE:B:D)
\end{equation*}
This exercise gives another illustration for how and why the reductive property motivates the method, by which we found inequality~(\ref{ourineq}). Setting $S_A = 0$ gives:
\begin{equation}
I(D:B:E) + I(D:BE:C) - I(D:B:CE) \geq 0 \label{redrewrite}
\end{equation}
The simplification occurs because monogamy $I(X:Y:Z) \geq 0$ reduces to $0 = 0$. Somewhat miraculously, region $B$ cancels out from this combination, leaving $I(C:D:E) \geq 0$.

\paragraph{Inequality~$i_8$} Under the symmetry of the inequality, $A$ can be mapped to $B$ or $C$ or $D$ but not to $E$ or $O$. The regions $E$ and $O$ can be exchanged, with accompanying transformations of the other regions. Therefore, up to symmetry, there are two distinct reductions: $S_A = 0$ and $S_E = 0$. The $S_A = 0$ one gives $I(B:C:E) \geq 0$. The $S_E = 0$ gives:
\begin{equation}
I(A:B:D) + I(A:C:D) \geq 0
\end{equation}

\paragraph{The new inequality} Regions $\{A,B,C\}$ are equivalent to one another up to symmetry (see Section~\ref{sec:symmetry}), as are $\{D,E,F,G,O\}$. Therefore, it is enough to consider reductions $S_A = 0$ and $S_D = 0$. Setting $S_A = 0$, we obtain:
\begin{equation}
S_{DEF} + S_{DEG} + S_{BCDE} + S_{BCDF} + S_{BCEG} \geq S_{BC} + S_{DE} + S_{DF} + S_{EG} + S_{BCDEF} + S_{BCDEG}
\end{equation}
Note that regions $BC$ always appear in combination. We now flip this duo with the purifier: $BC \leftrightarrow O$. After reordering terms, we obtain:
\begin{equation}
S_{DFO} + S_{FOG} + S_{OGE} + S_{GED} + S_{EDF} \geq S_{DF} + S_{FO} + S_{OG} + S_{GE} + S_{ED} + S_{DFOGE} 
\label{ourn5cyclic}
\end{equation}
This is the cyclic inequality on $N=5$. Inequality~(\ref{n5cyclicrewrite}) in the main text is equivalent to (\ref{ourn5cyclic}). 

The second reduction sets $S_D = 0$. We obtain:
\begin{align}
S_{ABE} + S_{ABF} + S_{ACE} + S_{ACF} \, + & \,  S_{BCE} + S_{BCF} + S_{AEF} + S_{BEF} + S_{CEF} \nonumber \\
& \geq \\
S_{AE} + S_{AF} + S_{BE} + S_{BF} + S_{CE} \, + & \, S_{CF} + S_{ABC} + S_{ABEF} + S_{ACEF} + S_{BCEF} \nonumber
\end{align}
This is inequality~$i_5$, as noted in the main text. 

\section{K-basis}
\label{sec:kbasis}

Reference~\cite{kbasis} defined a useful basis for the space of entropy assignments. It is constructed from graphs (see Section~\ref{sec:hyper}), which are perfect tensors on an even number of constituents. 

Let us denote the coefficients in the K-basis expansion of an entropy vector with $K_X$, where $X$ contains an even number of regions. For example, $K_{ABDF}$ is the coefficient of the perfect tensor on regions $\{A,B,D,F\}$ in the K-basis expansion of an entropy configuration. Reference~\cite{kbasis} proved the following statement: When we express a maximally tight entropy inequality---a facet of the entropy cone---in the K-basis, it must have only positive coefficients.

Written in the K-basis, inequality~(\ref{ourineq}) reads:
\begin{equation}
    \begin{split} \label{ourineqK}
 & K_{A B C D} + K_{A B C O} + K_{A B C E} + K_{A B C F} + K_{A B C G}\\
+ \, & K_{ADOE} + K_{AOEF} + K_{AEFG} + K_{AFGD} + K_{AGDO}\\
+ \, & K_{BDOE} + K_{BOEF} + K_{BEFG} + K_{BFGD} + K_{BGDO}\\
+ \, & K_{CDOE} + K_{COEF} + K_{CEFG} + K_{CFGD} + K_{CGDO}\\
+ \, & 3 K_{ABCDOE} + 3 K_{ABCOEF} + 3 K_{ABCEFG} + 3 K_{ABCFGD} + 3 K_{ABCGDO}\\
+ \, & 2 K_{ADOEFG}+ 2 K_{BDOEFG} + 2 K_{CDOEFG}\\
+ \, & K_{ABDOEF} + K_{ABOEFG} + K_{ABEFGD} + K_{ABFGDO} + K_{ABGDOE}\\
+ \, & K_{ACDOEF} + K_{ACOEFG} + K_{ACEFGD} + K_{ACFGDO} + K_{ACGDOE}\\
+ \, & K_{BCDOEF} + K_{BCOEFG} + K_{BCEFGD} + K_{BCFGDO} + K_{BCGDOE}\\
+ \, & 6 K_{A B C D E F G O} \geq 0
    \end{split}
\end{equation}
Our ordering in the subscripts and among the terms is not lexicographic. Instead, we write the terms so as to manifest the symmetry of the inequality. An $S_3$ permutes $\{A,B,C\}$ and a dihedral symmetry $D_5$ transforms regions $D$-$O$-$E$-$F$-$G$ like vertices of a regular pentagon. 

Going from inequality~(\ref{ourineq}) to expression~(\ref{ourineqK}) is done by an intricate change of basis in the 127-dimensional space of entropy assignments for $N=7$ named regions. That this change of basis returned only non-negative coefficients is a highly non-trivial sanity check for inequality~(\ref{ourineq}).

\end{document}